\documentclass[aps,prd,preprintnumbers,twocolumn,groupedaddress,nofootinbib]{revtex4}

\pdfoutput=1

\usepackage{graphicx}
\usepackage{latexsym}
\usepackage{amsfonts}
\usepackage{amssymb}
\usepackage{amsmath}
\usepackage{slashed}
\usepackage{array}
\usepackage{feynmp}
\usepackage[hypertex]{hyperref}
\usepackage{url}

\begin{document}
\title{Regenerating a Symmetry in Asymmetric Dark Matter}

\author{Matthew R.~Buckley$^{1}$ and Stefano Profumo$^2$}
\affiliation{$^1$Center for Particle Astrophysics, Fermi National Accelerator Laboratory, Batavia, IL 60510}
\affiliation{$^2$Department of Physics and Santa Cruz Institute for Particle Physics, University of California, 1156 High St., Santa Cruz, CA 95064, USA}

\preprint{FERMILAB-PUB-11-437-A}
\date{\today}

\begin{abstract}
\noindent Asymmetric dark matter theories generically allow for mass terms that lead to particle-antiparticle mixing. Over the age of the Universe, dark matter can thus oscillate from a purely asymmetric configuration into a symmetric mix of particles and antiparticles, allowing for pair-annihilation processes. Additionally, requiring efficient depletion of the primordial thermal (symmetric) component generically entails large annihilation rates. We show that unless some symmetry completely forbids dark matter particle-antiparticle mixing, asymmetric dark matter is effectively ruled out for a large range of masses, for almost any oscillation time-scale shorter than the age of the Universe.
\end{abstract}

\maketitle

The framework of asymmetric dark matter \cite{Nussinov:1985xr,Barr:1990ca} relates the existence and abundance of dark matter to the existence of the baryon asymmetry in the Universe.  Rather than take the coincidence that a thermal relic of the early Universe with weak-scale masses and couplings gives approximately the right amount of dark matter today (the ``WIMP miracle'') as a starting point for model-building, asymmetric dark matter models hinge on the fact that the matter density of baryons is not significantly different from the matter density of dark matter ($\rho_{\rm DM}/\rho_{\rm B} = 5.86$). If taken seriously, this coincidence implies that the asymmetry in the visible, baryonic sector is mirrored by an asymmetry in the dark sector. Thus, dark matter is not a thermal relic: rather, the presence of dark matter today is the result of the dark sector satisfying the Sakharov conditions for a baryogenesis-like process: containing CP violation, operators that violate dark matter number ($X$), and a departure from thermal equilibrium. Relating the matter densities of the two sectors implies that the generation of a dark asymmetry was related to the generation of a baryon asymmetry, perhaps through operators that violated both $B$ and $X$ simultaneously. Additionally, a characteristic requirement \cite{Buckley:2011kk} of all asymmetric dark matter models is a large dark matter annihilation cross section, as needed to eliminate the symmetric ({\it i.e.}~thermal) component.

Naively, it would seem that all asymmetric dark matter models would predict a complete lack of indirect detection signals,\footnote{A possible exception might be detecting the effect of gamma radiation from cosmic-ray scattering off of galactic dark matter, albeit the expected signal is generically faint \cite{Profumo:2011jt}} as the dark matter would be composed primarily of the dark matter particle $\psi$, with a highly suppressed component of the $\bar{\psi}$ antiparticle.\footnote{Throughout the paper we will use the notation of a fermion ($\psi/\bar{\psi}$) for dark matter. Unless otherwise noted, though, our arguments will apply equally to scalar dark matter ($\varphi/\varphi^*$).} However, as we shall show, this is not generically true. 

A defining feature of dark matter is that it is uncharged under electromagnetism and the strong interaction. Thus, like the neutrino, dark matter has Standard Model gauge charges allowing $\Delta X = 2$ mass terms:
\begin{eqnarray}
{\cal L}_{\rm fermion} & \supset & m_D \psi \bar{\psi} + m_M (\psi\psi+\bar{\psi}\bar{\psi}) \label{eq:fermion} \\
{\cal L}_{\rm scalar} & \supset & m_D^2 \varphi \varphi^* + m_M^2 (\varphi\varphi+\varphi^*\varphi^*). \label{eq:scalar}
\end{eqnarray}
As a result, even if an $X$-asymmetry is generated at high scales, the particle $\psi$ will oscillate back into $\bar{\psi}$ over timescales on the order of $\Delta m = 2 m_M$ for fermionic dark matter ($\Delta m = m^2_M/m_D$ for scalars). As we shall show in detail, significant bounds exist on asymmetric dark matter models with oscillation timescales shorter than the age of the Universe ($\tau_{\rm Universe} \sim 10^{17}~\mbox{s} \sim 10^{41}~\mbox{GeV}^{-1}$). Assuming a ``seesaw mechanism'' similar to that proposed for neutrino masses \cite{seesaw}, 
requiring $\Delta m \ll \tau_{\rm Universe}^{-1}$ places stringent limits on new physics up to a scale beyond $M_{\rm Planck}$. Therefore, barring some compelling model-building reason to forbid $\Delta X = 2$ mass terms (for example, charging the dark matter under new gauge symmetries, as often found in technicolor or confining dark matter models \cite{Nussinov:1985xr}), we argue that indirect detection signals are a generic expectation of asymmetric dark models. Interestingly, if the thermally averaged pair-annihilation cross section times relative velocity of dark matter particles and antiparticles into Standard Model final states $\langle\sigma v_{\rm rel}\rangle_{\psi\bar\psi\to{\rm S.M.}}\equiv\langle\sigma v\rangle$ is large enough to wipe out the symmetric relic component and let the asymmetric component dominate, related indirect signals are expected to be comparitively large and within the sensitivity of current instruments such as gamma-ray telescopes. 

\begin{figure*}[ht]
\mbox{\includegraphics[width=0.48\textwidth,clip]{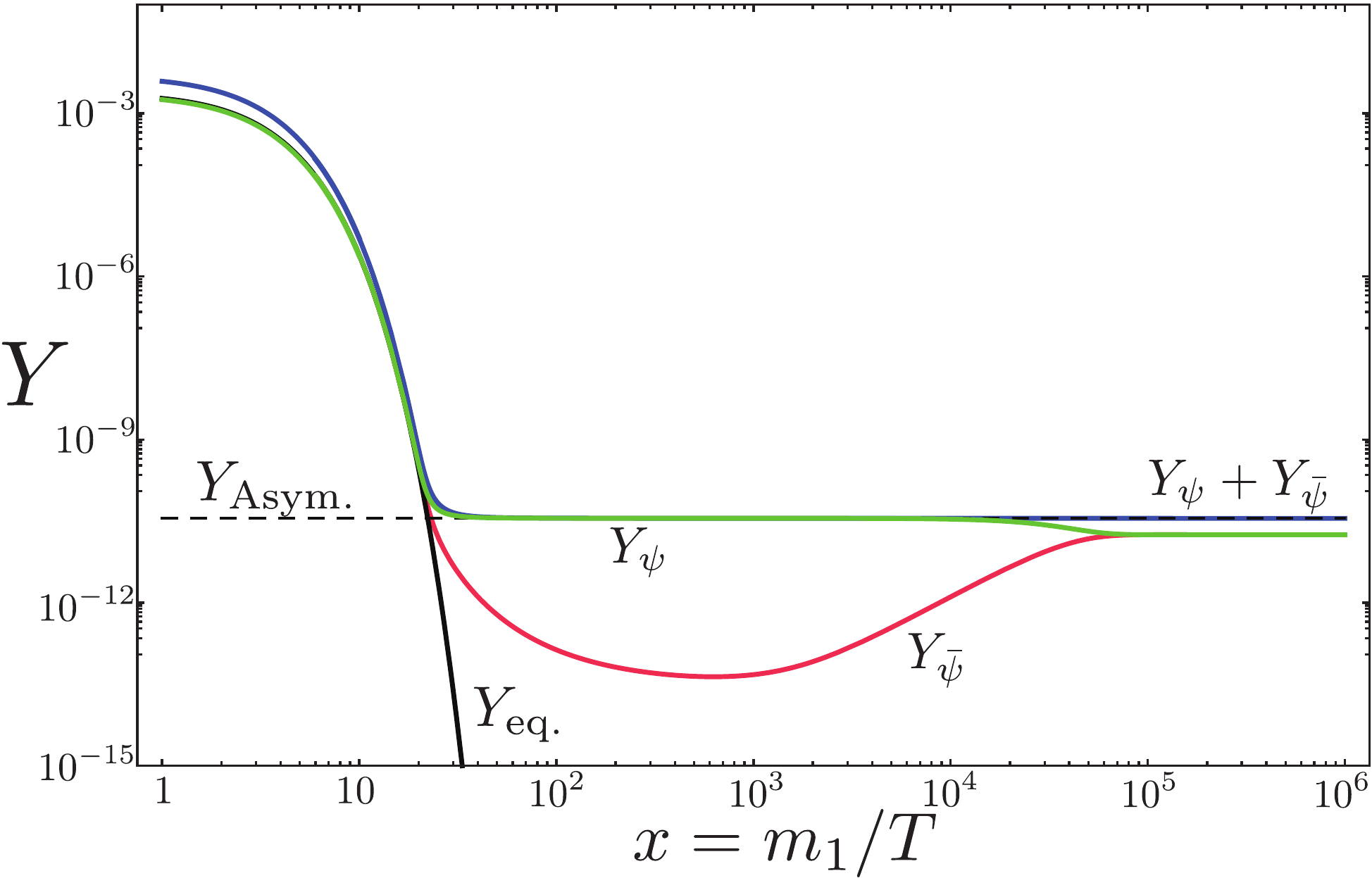}\qquad\includegraphics[width=0.48\textwidth,clip]{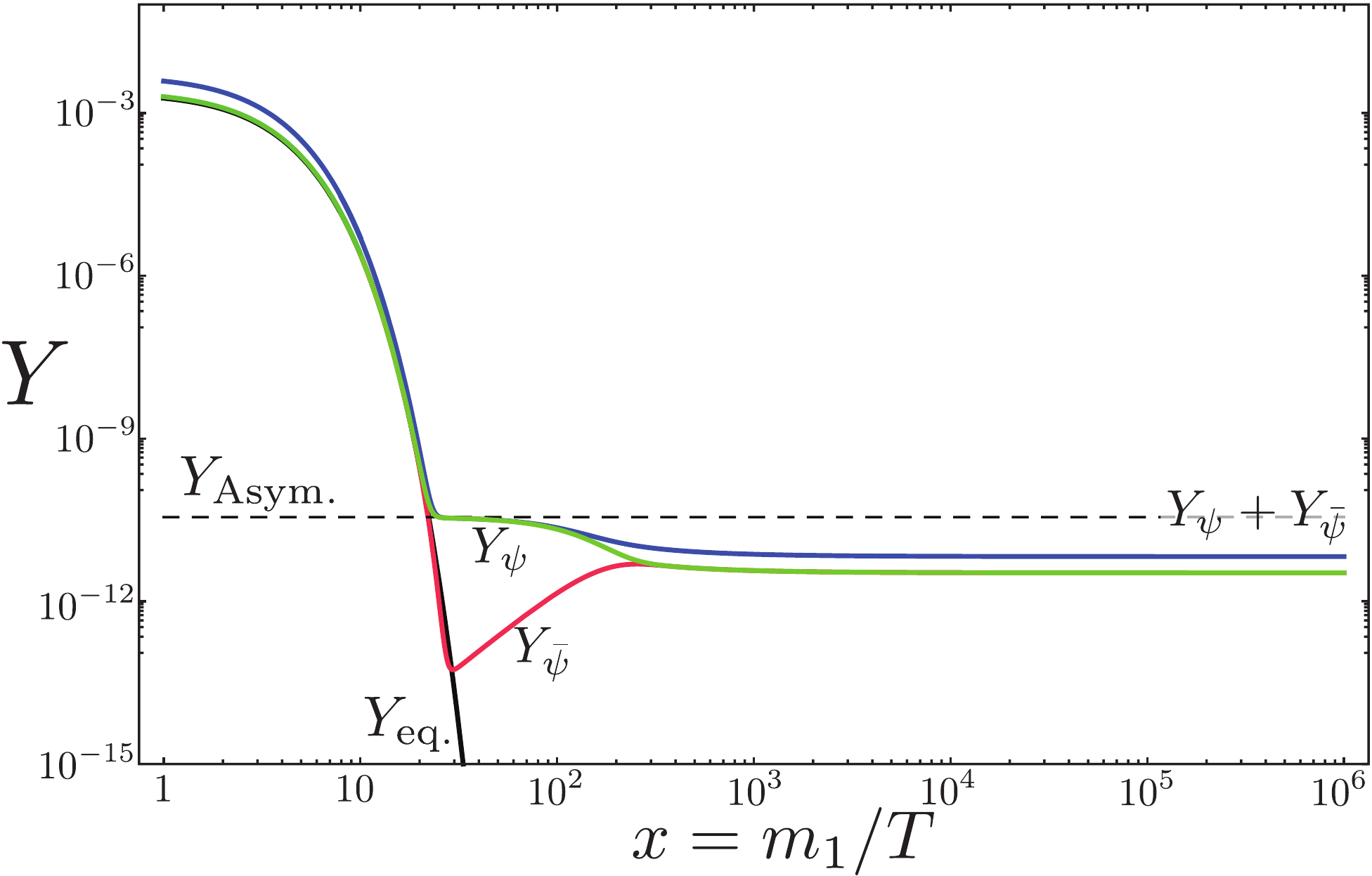}}
\caption{The number density over entropy density $Y\equiv n/s$ as a function of $x\equiv m_1/T$, for $m_1=10$ GeV, $\langle\sigma v\rangle=1.5\times 10^{-25}$~cm$^3$/s and $\Delta m=10^{-25}$~GeV (left) and $m_1 = 1$~TeV, $\langle \sigma v \rangle = 10^{-24}$~cm$^3$/s, and $\Delta m = 8 \times 10^{-21}$~GeV (right).\label{fig:annihilation}}
\end{figure*}

With this motivation, we explore the allowed regions of dark matter-Standard Model interaction cross-section $\langle \sigma v\rangle$ vs.~oscillation timescale $\tau$. For specificity, we assume that dark matter annihilates into $b\bar{b}$ final states.\footnote{Though most of the relevant bounds are not very sensitive to this choice, we note that the BBN constraints do depend on whether the final states are hadronic, and annihilation into $\nu\bar\nu$ has significantly relaxed limits. We defer detailed consideration of these cases at this time.} Working with the Lagrangians of Eqs.~\eqref{eq:fermion}-\eqref{eq:scalar}, we designate the lighter mass eigenstate as $m_1$ and the heavier as $m_2$. Starting with a pure state of a non-relativistic $|\psi\rangle$ particle, the probability of finding $|\bar{\psi}\rangle$ after time $t$ is
\begin{equation}
P(|\psi\rangle \to |\bar{\psi}\rangle) = \sin^2\left( \frac{\Delta m t}{2} \right). \label{eq:probability}
\end{equation}

\begin{figure*}[ht]
\mbox{\includegraphics[width=0.48\textwidth,clip]{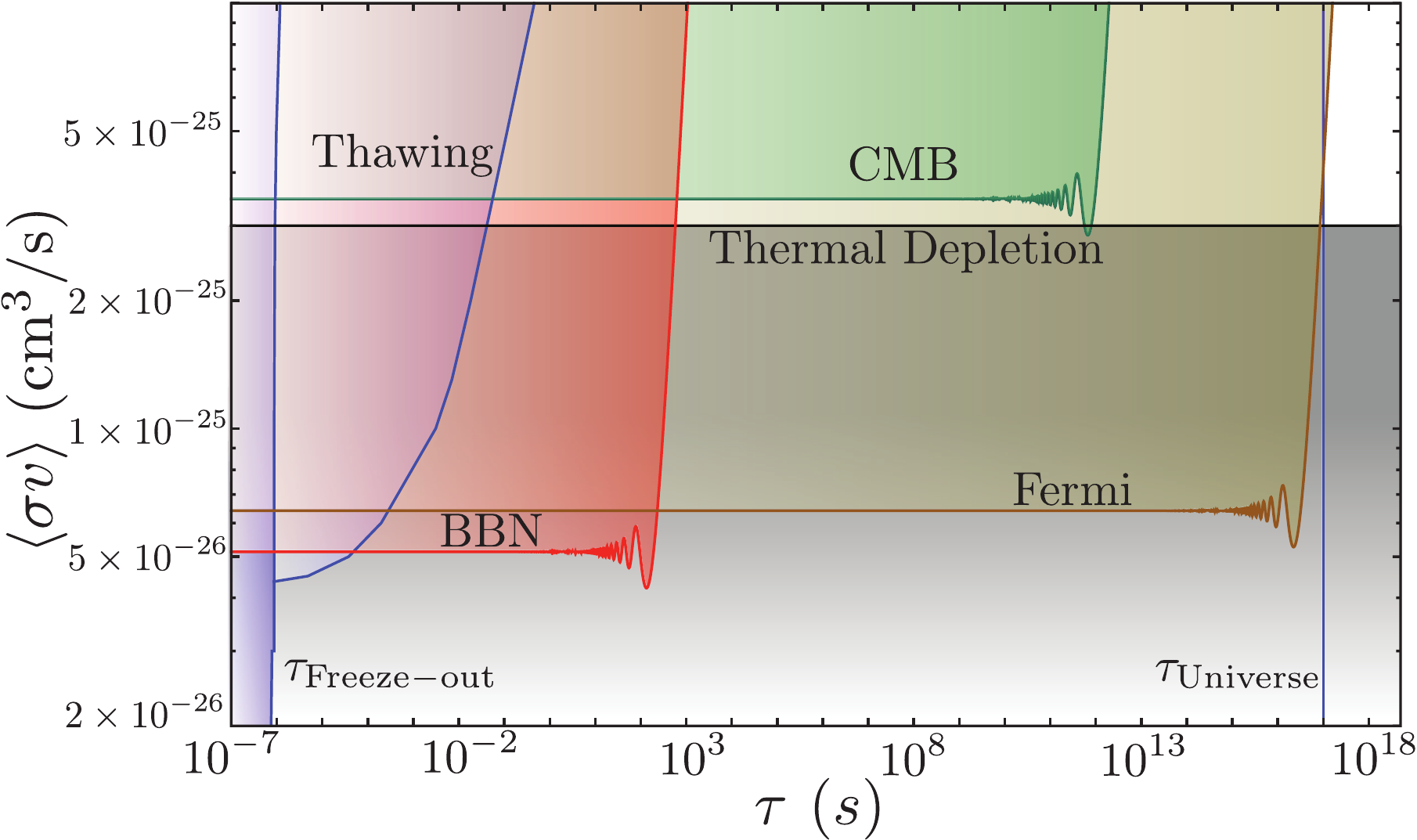}\qquad\includegraphics[width=0.48\textwidth,clip]{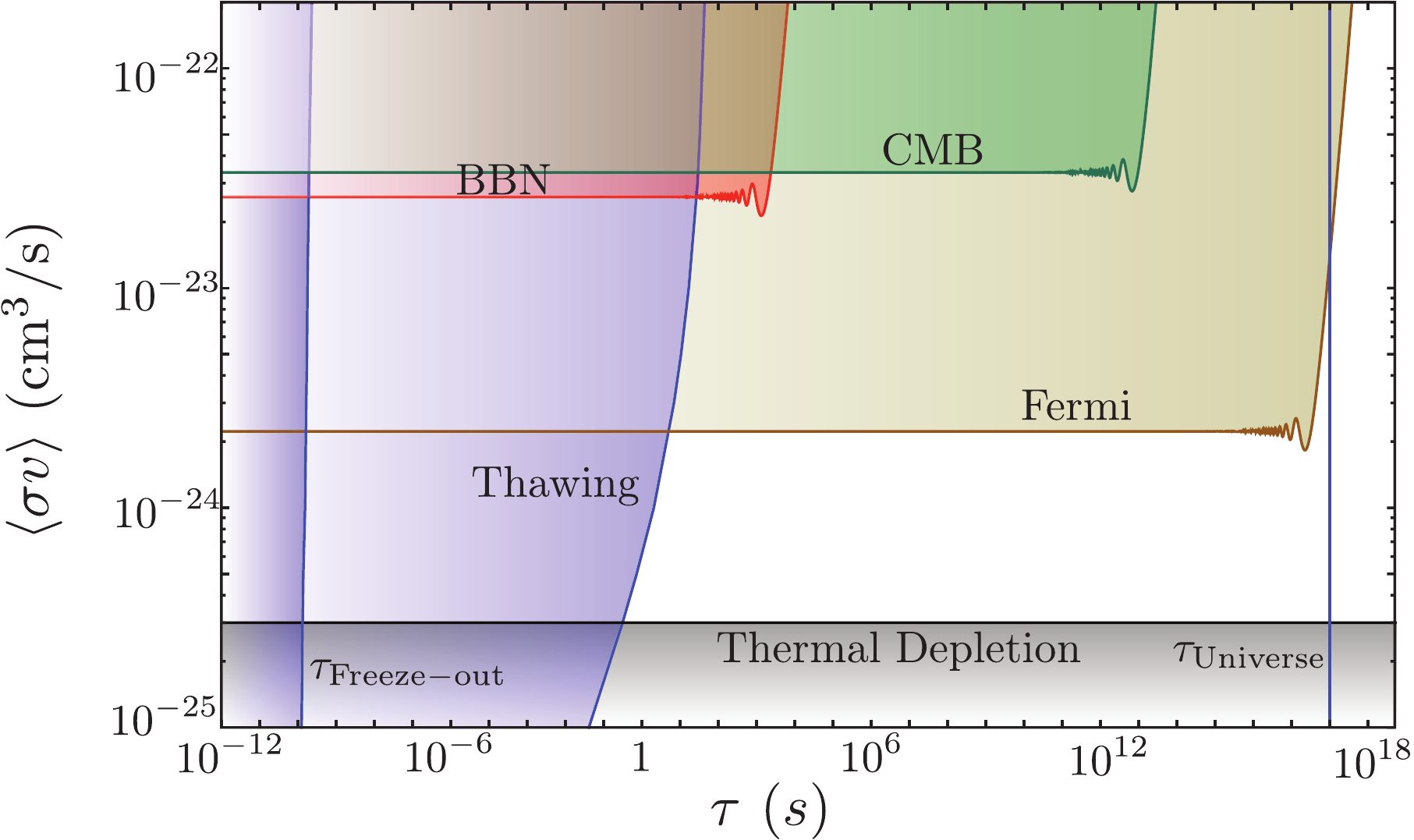}}
\caption{Constraints on the $(\tau,\langle\sigma v\rangle)$ parameter space for $m_1 = 10$~GeV (left) and 1000~GeV (right); the shaded regions are excluded, the white regions allowed -- see the text for details.\label{fig:limits}}
\end{figure*}

We choose two benchmark masses for dark matter, $m_1=10$~GeV and 1000~GeV. We note that light dark matter is a common result of asymmetric models, though TeV-scale dark matter is also possible through Boltzmann suppression at the time when the operator that allows transfer of $X$ into $B$ decouples \cite{Buckley:2010ui}. The combined limits will be described below, and are summarized in the plots of Fig.~\ref{fig:limits}.

The evolution of the number density $n_i$ of the dark matter particles $\psi$ and antiparticles $\bar \psi$ is set by two coupled Boltzmann equations. As is customary, we work in the variables $x \equiv m_1/T$ and $Y_i \equiv n_i/s$, where $s$ is the entropy density. Neglecting processes like $\psi \leftrightarrow \bar{\psi}$ conversion via scattering off the cosmic thermal background, for $i,j=\psi,\bar\psi$, we find (see {\it e.g.}~Ref.~\cite{Gondolo:1990dk})
\begin{eqnarray}
\frac{{\rm d}Y_i}{{\rm d}x}=-\langle\sigma v \rangle\sqrt{\frac{\pi}{45G}}\frac{m_1\ g_*^{1/2}}{x^2}\left(Y_iY_j-Y_{\rm eq}^2\right)&&\\
\nonumber-\Gamma_{ij}\frac{g_*^{1/2}}{h_{\rm eff}}\sqrt{\frac{45}{4\pi^3 G}}\frac{x}{m_1^2}\left(Y_i-Y_j\right).&&
\end{eqnarray}
Here,
\begin{equation}
g_*^{1/2}\equiv\frac{h_{\rm eff}}{g_{\rm eff}}\left(1+\frac{T}{3h_{\rm eff}}\frac{{\rm d}h_{\rm eff}}{{\rm d}T}\right),
\end{equation}
and $h_{\rm eff}$ and $g_{\rm eff}$ are the effective energy and entropy density degrees of freedom \cite{Gondolo:1990dk}. 
$\Gamma_{ij}$ is the rate of $\psi \to \bar{\psi}$ conversion:
\begin{equation}
\Gamma_{ij}=\Gamma_{ji}\equiv\Gamma=\Delta m \equiv \tau^{-1}.
\end{equation}
In the large-$x$ regime, defined as $x\gg x_{\rm freeze-out}\equiv x_{\rm f.o.}$, the system of differential equations simplifies to
\begin{eqnarray}
\label{eq:bol1}\frac{\rm d}{{\rm d}x}\left(Y_\psi+Y_{\bar \psi}\right)&=&0\\
\label{eq:bol2}\frac{\rm d}{{\rm d}x}\left(Y_\psi-Y_{\bar \psi}\right)=\frac{{\rm d}\delta}{{\rm d}x}&=&-2\Delta m \frac{g_*^{1/2}}{h_{\rm eff}}\sqrt{\frac{45}{4\pi^3 G}}\frac{x}{m_1^2} \delta.
\end{eqnarray}
This implies that there is no significant $\bar\psi$ regeneration when the Universe is at a temperature $T$ as long as
\begin{equation}
\Delta m\lesssim \frac{h_{\rm eff}}{g_*^{1/2}}\sqrt{\frac{4\pi^3}{45}}\ \frac{T^2}{m_{\rm Pl}}.
\end{equation}

To qualify as an asymmetric model, we (somewhat arbitrarily) require that 90\% of the dark matter density originates from the asymmetric component, rather than from symmetric $\psi$ and $\bar\psi$ arising from thermal freeze-out, which must  be thus smaller than 10\% of the observed cosmological dark matter density. This immediately places a $\tau$-independent lower bound on the annihilation rate (labeled ``Thermal Depletion'' in Fig.~\ref{fig:limits})
\begin{equation}
\langle \sigma v\rangle \gtrsim  10\times\left(3\times 10^{-26}~\mbox{cm$^3$/s}\right), \label{eq:freezeout}
\end{equation}
where, for simplicity, we assume an $s$-wave dominated pair annihilation cross section.

Depending upon the hierarchy between $\tau$ and the freeze-out time, one generically has four cases:
\begin{enumerate}
\item $\tau\sim1/\Delta m\ll t_{\rm f.o.}$. Here $\psi-\bar\psi$ mixing happens before freeze-out, the two species are coupled before and throughout freeze-out, and a relic density $\Omega_{\rm th}\sim 3\times 10^{-27}/\langle \sigma v\rangle$ of both $\psi$ and $\bar\psi$ is leftover, independent of the initial asymmetric component. Therefore, according to our definition, the dark matter model is not asymmetric: the final abundance is set not by the size of the $\psi$ asymmetry, but by the thermal cross section.\item $\tau\sim1/\Delta m\gg t_{\rm f.o.}$, and residual annihilations at $t\gtrsim\tau$ do not substantially modify the total $\psi+\bar\psi$ number density (to be quantitative, we define a substantial modification as a 10\% effect). In this case, the relic density of $\bar \psi$ oscillates up to (half of) the initial asymmetric density, and so there are indirect detection constraints from $\psi\bar\psi$ pair annihilation. This corresponds to the regime of validity of the pair of Eqs.~\eqref{eq:bol1}-\eqref{eq:bol2}. We illustrate this case with the left panel of Fig.~\ref{fig:annihilation}.
\item $\tau\sim1/\Delta m \gtrsim t_{\rm f.o.}$, and residual annihilations substantially (at a level more than 10\%) modify the total $\psi+\bar\psi$ number density. The frozen asymmetric component ``{\em thaws}'' and yields a (suppressed) abundance of $\psi$ and $\bar\psi$ pairs. This typically happens for a large $\langle \sigma v\rangle$, and for a short $\tau$ that is not too much larger than $t_{\rm f.o.}$. This case is illustrated in the right panel of Fig.~\ref{fig:annihilation}.
\item $\tau\gg \tau_{\rm Universe}$: this corresponds to $\Delta m\ll 10^{-41}$ GeV. In this case there effectively are no oscillations; the final relic density corresponds to the asymmetric component, as long as the latter is much larger than the residual symmetric one; there are no residual pair annihilations, and dark matter is indeed asymmetric.
\end{enumerate}

The cross-over between cases 2 and 3 depends non-trivially on the annihilation rate and on the $\psi-\bar\psi$ oscillation time-scale. We illustrate this, for $m_1=10$ GeV and 1000 GeV, in the left and right panels, respectively, of Fig.~\ref{fig:limits}, with the line labeled ``Thawing.'' The region to the left of that line has a final dark matter density smaller than (90\% of) the asymmetric component; the region thus requires a relatively larger asymmetric component than in the case where annihilations do not affect the dark matter $\psi+\bar\psi$ number density. The shape of the region is what expected: the later the $\psi\bar\psi$ oscillation occurs, the larger the pair-annihilation cross section needed to affect the total number density. As clear from Fig.~\ref{fig:annihilation}, right, across the ``Thawing'' region, $Y_\psi\simeq Y_\psi$ and annihilation processes occur in the early universe and today, as is also the case for cases 1 and 2. We now discuss in detail the remaining  constraints on oscillating asymmetric dark matter from $\psi-\bar\psi$ annihilation.

In thermal dark matter models, over-production of $^6$Li during Big Bang Nucleosynthesis (BBN) can be used to place an upper limit on the annihilation cross section, see {\it e.g.}~Ref.~\cite{Jedamzik:2004ip}. These limits can be applied to an oscillating asymmetric scenario by using an effective cross section, in which the fundamental cross section $\langle \sigma v\rangle$ is re-weighed by the average amount of $\bar{\psi}$ that exists during BBN. In general, $\langle \sigma v\rangle$ in an asymmetric model with $\tau \equiv \Delta m^{-1}$ is made equivalent to a value of $\langle \sigma v\rangle_{\rm sym}$ for symmetric dark matter at time $t$ by equating $4 n_{\rm sym.}^2 \langle \sigma v \rangle_{\rm sym.} = n_\psi(t) n_{\bar{\psi}}(t) \langle \sigma v\rangle$ (the 4 accounts for the Majorana nature of neutralinos). Integrating the probability Eq.~\eqref{eq:probability} from time zero to $t$ and normalizing gives $n_{\bar{\psi}}$ in terms of $n_\psi$, and assuming $n_\psi(0) = n_{\rm sym.}$, we find
\begin{equation}
\langle \sigma v\rangle = \frac{4 \langle \sigma v \rangle_{\rm sym.}}{1-\frac{\sin t/\tau }{t/\tau}}. \label{eq:asymaverage}
\end{equation}

From Ref.~\cite{Jedamzik:2004ip}, the upper bound on $\langle \sigma v\rangle_{\rm sym.}$ for a 10(1000)~GeV symmetric dark matter particle annihilating into $b\bar{b}$ is $4.3 \times 10^{-26}(6.6 \times 10^{-24})$~cm$^3$/s. The derived upper bounds on $\langle \sigma v\rangle$ as a function of $\tau$ for the two benchmark models are shown in Fig.~\ref{fig:limits}, assuming $t=\tau_{\rm BBN} \equiv 600$~s. We neglect possible additional constraints from depletion of primordial abundances due to annihilations at times greater than $\tau_{\rm BBN}$.

Another constraint on the annihilation cross section is derived from data on the Cosmic Microwave Background (CMB). The temperature and angular power spectrum are sensitive to energy injection from dark matter annihilation into Standard Model final states, which reionizes the photon-baryon plasma. The current constraints from WMAP data are analyzed in Ref.~\cite{Slatyer:2009yq}, from where we take the upper bound on symmetric dark matter to be
\begin{equation}
\langle \sigma v \rangle_{\rm sym.} < \frac{3.6 \times 10^{-24}~\mbox{cm$^3$/s}}{f} \left( \frac{m_1}{1~\mbox{TeV}}\right)
\end{equation}
For annihilation into $b\bar{b}$ at $z=2500$, the parameter $f$ is estimated to be $\sim 0.41$. As for the BBN bounds, the limit for asymmetric models must be reweighted by Eq.~\eqref{eq:asymaverage}; the limits are shown in Fig.~\ref{fig:limits}, where we have assumed $t= \tau_{\rm CMB} \equiv 10^{13}$~s.

Finally, bounds on annihilation of dark matter in the Universe today can be derived by indirect detection searches. While $\langle\sigma v\rangle$ can be constrained in a variety of ways from cosmic-ray and gamma-ray observations (for a theorists' perspective see {\it e.g.}~Ref.~\cite{Cirelli:2010xx}), for definiteness we choose to consider here the recent limits obtained from stacked observations of nearby dwarf spheroidal galaxies with the Fermi Gamma-Ray Space Telescope \cite{collaboration:2011wa}. 
For our two benchmark models, again assuming annihilation into $b\bar{b}$, the limits are $\langle \sigma v\rangle_{\rm sym.} < 1.6 \times 10^{-26}(5.5 \times 10^{-25})$~cm$^3$/s for 10(1000)~GeV. We plot the derived  bounds on asymmetric dark matter models on the $(\tau,\langle\sigma v\rangle)$ parameter space in Fig.~\ref{fig:limits}, setting $t = \tau_{\rm Universe} \equiv 10^{17}$~s. 

Inspection of the combined bounds illustrates that asymmetric dark matter models with light masses (on the order 10 GeV) are essentially ruled out, unless an extremely small $\Delta m\ll 10^{-41}$ GeV, corresponding to $\tau\gtrsim \tau_{\rm Universe}$, is produced by $\Delta X=2$ operators. A small window opens up for masses in the hundreds of GeV range -- less relevant to explain recent anomalies in direct detection experiments (see {\it e.g.}~\cite{Hooper:2011hd}) -- as can be inferred from the right panel, where $m_1=1$ TeV.
Altering the final states and masses, or postulating dominant $p$-wave annihilation in $\langle\sigma v\rangle$ does not significantly change these results or alter our conclusions.

With such stringent bounds, equivalent to requiring that $\Delta m \lesssim 10^{-41}$~GeV, it would seem that there must exist some symmetry that is not even violated by Planck-scale suppressed operators. Since global symmetries are generically assumed to be violated by such gravitational operators \cite{globalplanck}, it seems possible that $\Delta X =2$ operators must be forbidden by gauge symmetries (or a discrete symmetry, perhaps originating from some gauge symmetry at high energies), if oscillating dark matter is to be avoided and if dark matter truly be asymmetric without conflicting with observations. 

\begin{acknowledgments}
We would like to thank P.~Fox, G.~Kribs and D.~Hooper for discussions. MRB is supported by the US Department of Energy. Fermilab is operated by Fermi Research Alliance, LLC under Contract No.~\protect{DE-AC02-07CH11359} with the US~Department of Energy. SP is partly supported by an Outstanding Junior Investigator Award from the US Department of Energy and by Contract DE-FG02-04ER41268, and by NSF Grant PHY-0757911.  The authors gratefully acknowledge the hospitality of the Aspen Center for Physics, supported by NSF grant 1066293, where this work was started.
\end{acknowledgments}

\end{document}